# Direct observation of distinct minibands in moiré superlattices

Saien Xie[1,2,3]*, Brendan D. Faeth[1], Yanhao Tang[4], Lizhong Li[4], Christopher T. Parzyck[1], Debanjan Chowdhury[1], Ya-Hui Zhang[5], Christopher Jozwiak[6], Aaron Bostwick[6], Eli Rotenberg[6], Jie Shan[1,3,4], Kin Fai Mak[1,3,4], Kyle M. Shen[1,3]*

[1] Department of Physics, Laboratory of Atomic and Solid State Physics, Cornell University, Ithaca, NY, USA

[2] Department of Materials Science and Engineering, Cornell University, Ithaca, NY, USA

[3] Kavli Institute at Cornell for Nanoscale Science, Ithaca, NY, USA

[4] School of Applied and Engineering Physics, Cornell University, Ithaca, NY, USA

[5] Department of Physics, Harvard University, Cambridge, MA, USA

[6] Advanced Light Source, E. O. Lawrence Berkeley National Laboratory, Berkeley, CA, USA

* Correspondence to: sx68@cornell.edu and kmshen@cornell.edu

**Moiré superlattices comprised of stacked two-dimensional materials present a versatile platform for engineering and investigating new emergent quantum states of matter[1–16]. At present, the vast majority of investigated systems have long moiré wavelengths[17–20], but investigating these effects at shorter, incommensurate wavelengths, and at higher energy scales, remains a challenge. Here, we employ angle-resolved photoemission spectroscopy (ARPES) with sub-micron spatial resolution to investigate a series of different moiré superlattices which span a wide range of wavelengths, from a short moiré wavelength of 0.5 nm for a graphene/WSe$_2$ (g/WSe$_2$) heterostructure, to a much longer wavelength of 8 nm for a WS$_2$/WSe$_2$ heterostructure. We observe the formation of minibands with distinct**

**dispersions formed by the moiré potential in both systems. Finally, we discover that the WS$_2$/WSe$_2$ heterostructure can imprint a surprisingly large moiré potential on a third, separate layer of graphene (g/WS$_2$/WSe$_2$), suggesting a new avenue for engineering moiré superlattices in two-dimensional materials.**

In principle, virtually all moiré superlattices are only quasiperiodic and do not formally obey Bloch's theorem, so traditional concepts such as the Brillouin zone and band structure can become problematic. At present, one of the more successful approaches for explaining and understanding the properties of moiré superlattices is to treat the incommensurability between the different layers in a continuum limit as a smoothly varying function of position[17–20], an approximation well suited for long moiré wavelengths, appropriate for systems such as "magic angle" twisted bilayer graphene[17] (13 nm), trilayer graphene on boron nitride[16] (15 nm), and aligned WS$_2$/WSe$_2$ bilayers[19,20] (8 nm). On the other hand, the properties of moiré superlattices at shorter, incommensurate wavelengths, or at higher energy scales, remains a challenge to both theory and experiment. Experimental probes such as transport and optical spectroscopy are generally most sensitive only to the lowest lying electronic states, typically near the K point for graphene and monolayer transition metal dichalcogenides (TMDs). Therefore, many open questions remain about moiré superlattices at shorter length scales or at higher energy scales which are inaccessible to transport measurements. For instance, do minibands still form in quasiperiodic systems when the moiré wavelength is comparable to the lattice constants of the constituent materials? What is the behavior of minibands at higher energies, away from the band edges? Is it possible to imprint a moiré superpotential comprised from a bilayer onto a third, separate layer?

Here, we employ angle-resolved photoemission spectroscopy (ARPES) with sub-micron spatial resolution to address these questions by investigating a series of different moiré superlattices. In a

graphene/WSe$_2$ (g/WSe$_2$) heterostructure with a very short moiré wavelength of 0.5 nm (Fig. 1a), a regime difficult to treat from a theoretical basis[19,20], we observe moiré minibands which exhibit a wholly distinct dispersion from either the graphene or WSe$_2$ low-energy bands. In an aligned WS$_2$/WSe$_2$ heterostructure with a much longer moiré wavelength of 8 nm (Fig. 1b), we observe distinct minibands near Γ, with an unexpectedly large energy splitting of 220 meV between the first and second band, much larger than the energy splitting of ~15 meV at the valence band maximum[21] (K point). We also reveal that in a trilayer heterostructure of graphene/WS$_2$/WSe$_2$ (g/WS$_2$/WSe$_2$), the WS$_2$/WSe$_2$ heterostructure can imprint a surprisingly large moiré potential on the graphene layer, presenting a new avenue for future flat-band engineering for studying effect of strong correlations in quasiperiodic structures.

In Figures 1c and d, we show an optical micrograph and schematic of a moiré superlattice sample investigated in this study. Due to the different sizes of the constituent flakes, different kinds of superlattices such as g/WSe$_2$ (Region 1) and g/WS$_2$/WSe$_2$ (Region 2) are simultaneously formed in this sample. Because of the small size of the beam spot (< 1 μm$^2$), each region can be identified by their electronic structure and investigated independently. For example, in Figure 1e, ARPES measurements in Regions 1 and 2 are shown. In Region 1, both the WSe$_2$ and graphene bands are evident, with clear gaps where the WSe$_2$ and graphene bands intersect indicating hybridization between the electronic states in the two layers. In Region 2, both the WS$_2$ and WSe$_2$ valence bands are visible, in addition to the graphene Dirac cones, which form multiple replicas arising from interactions with the WS$_2$/WSe$_2$ moiré potential, as will be described later in the text. By selectively plotting the intensity from either the WS$_2$ bands (in red; only present in the region with the WS$_2$/WSe$_2$ moiré, due to the placement of the WS$_2$ flake shown in Fig. 1c) or the WSe$_2$ bands (in blue), we can construct a spatial map of the ARPES intensity (Fig. 1f) corresponding to the regions comprised of g/WS$_2$/WSe$_2$ or g/WSe$_2$, which are in good agreement with the optical images. In this study, we investigate two separate, independent samples with different twist angles, one comprised solely of g/WSe$_2$ (with a twist angle of 27.3º, shown in Figure 2), and one comprised of g/WS$_2$/WSe$_2$ (where the WS$_2$ and WSe$_2$ are aligned, with a twist angle of 10º relative to graphene, shown in Figures 1, 3, and 4).

We begin by discussing moiré superlattices in the limit of short, incommensurate wavelengths (i.e., large wavevectors) in g/WSe$_2$ heterostructures, where the graphene is rotated by 27.3º relative to the WSe$_2$ layer (Brillouin zone schematically shown in Fig. 2a). Here, the moiré wavelength is approximately 5.1 Å, which is less than 2 WSe$_2$ unit cells ($a$ = 3.28 Å). This case is difficult to treat theoretically, since the conventional approach of using a continuum approximation[17–20] to treat the interlayer moiré potential is problematic, as the local interactions will vary dramatically and not smoothly from site to site. In Figure 2b, we show intensity maps at various binding energies as a function of in-plane momentum for the g/WSe$_2$ heterostructure. At the Fermi energy ($E_F$), sharp Dirac points are visible only at the graphene K points ($K_G$), indicating that the graphene is at charge neutrality with minimal charge transfer from the WSe$_2$ (sheet carrier density estimated to be less than $1.66 \times 10^{10}$ cm$^{-2}$). With increasing binding energy, the top of the WSe$_2$ valence band at the WSe$_2$ K point ($K_{WSe2}$) becomes apparent as a pair of spin-orbit split pockets, while the graphene Dirac cone opens into the expected "horseshoe"-shaped arc at $K_G$. These respective features are consistent with the expected electronic structure of the individual, non-interacting graphene and WSe$_2$ monolayers, as seen by the close agreement between the ARPES map and DFT results in the bottom panel of Fig. 2b. In addition to those features, we also observe a sextet of distinct minibands that arise from the moiré superlattice which reach their maxima at wavevectors of $K_{WSe2}$ + $G_{moiré}$ (Figure 2c, where $G_{moiré}$ is the moiré wavevector) and at a binding energy of 0.86 ± 0.02 eV (Figure 2d). In addition, the extracted dispersion of the moiré minibands is shown in Fig. 2e and compared to the WSe$_2$ valence band from $\Gamma$ to K$_{WSe2}$, as well as the graphene Dirac cone. The fact that the dispersion and effective mass of the moiré minibands (m* = 0.14 m$_0$, m$_0$ being the free electron mass) is substantially different from both the WSe$_2$ valence bands (m* = 1.42 m$_0$ at $\Gamma$ and m* = 0.68 m$_0$ at K$_{WSe2}$) or graphene Dirac cone (v$_F$ = 1.13×10$^6$ m s$^{-1}$) underscores that these minibands are new, emergent bands which arise from hybridization between the graphene and WSe$_2$ layers, as opposed simply to replicas of the WSe$_2$ valence bands at K$_{WSe2}$ that are folded back to the K$_{WSe2}$ + G$_{moiré}$ points.

The formation of such emergent minibands in the g/WSe$_2$ moiré superlattice is remarkable, particularly given work on twisted bilayer graphene which suggest that the strong renormalization of the minibands occurs only at very long wavelengths. The observed g/WSe$_2$ miniband shows a distinct dispersion, making the strong moiré potential between the graphene and WSe$_2$ layers unambiguous, in contrast to many other observations of minibands which are folded replicas of the original bands[22] or replica bands due to Umklapp scattering processes[23]. This strong potential is surprising given its very short periodicity, more than an order of magnitude smaller than those in magic-angle graphene (13 nm) and aligned WS$_2$/WSe$_2$ (8 nm), since in those systems a short moiré periodicity (arising from a large twist angle) results in an effective decoupling of the two monolayers[11,24]. Finally, our measurements probe minibands away from the valence band maximum (at K$_{WSe2}$), a region inaccessible to prevailing optical or transport measurements. These results suggest that moiré superlattice effects are richer and more complex than previously expected in heterobilayers, motivating a generalized understanding of quasiperiodic systems with arbitrary moiré wavelengths where the existing models are invalid.

Even in longer wavelength moiré superlattices, there remain challenges to our generalized understanding of their electronic structure and properties. For example, in the aligned $WS_2/WSe_2$ superlattice, which has a moiré wavelength of 8 nm, the existence of moiré minibands and excitons has been the subject of extensive investigation by optical and transport measurements [9–14], which probes these states around the valence band maximum ($K_{TMD}$). However, their behavior throughout momentum space remains an open question. In Fig. 3b, we show ARPES spectra from g/$WS_2$/$WSe_2$ (Brillouin zone schematically shown in Fig. 3a); at the $\Gamma$ point, the relevant graphene bands are far away in energy (~ 3 eV), so the relevant electronic states belong only to $WS_2$/$WSe_2$. In a non-interacting scenario, $WSe_2$ and $WS_2$ each would individually contribute only a single band, with nearly two-fold spin degeneracy, at $\Gamma$. However, in Fig. 3b and 3c, three bands are clearly observed at $\Gamma$, at binding energies of 1.16 eV, 1.38 eV and 1.80 eV, respectively. This large splitting of 220 ± 5 meV between the lowest-lying moiré miniband and the middle band at $\Gamma$ is striking, given that the splitting of the moiré minibands at the $K_{TMD}$ point is only ~ 15 meV[21]. The reason for this discrepancy remains an open question, but may stem from the different nature of the transition metal $d$ bands, which have a primarily in-plane orbital character at $K_{TMD}$, but have substantially more out-of-plane character at $\Gamma$, and might thereby facilitate stronger interlayer hybridization. We note that in a previous report of $MoSe_2$/$WSe_2$, a comparatively large splitting of 200 meV was observed at $\Gamma$ between the $WSe_2$ valence band and the hybridized heterobilayer band, and was hypothesized to arise from commensuration between the closely lattice-matched $MoSe_2$ and $WSe_2$ layers ($\Delta a/a$ ~ 0.3%)[24]. However, in the case of $WS_2$ and $WSe_2$, the lattice mismatch is much larger ($\Delta a/a$ ~ 4%), making it unlikely that the layers could be commensurate. Another signature of the impact of the moiré superlattice is that the bands around $\Gamma$ are significantly flatter than expected from the individual $WSe_2$ or $WS_2$ monolayers, potentially due to zone folding from the small moiré Brillouin zone. For instance, the effective mass for the middle band at Γ in g/WS$_2$/WSe$_2$ is approximately m* = 4.1 m$_0$, as opposed to only m* = 1.4 m$_0$ in the g/WSe$_2$. This large splitting at Γ suggests that moiré superlattice effects can differ by as much as an order of magnitude, depending on the location in the Brillouin zone. Given the large energy splittings observed here (i.e., hundreds of meV), it may be possible to construct moiré superlattices with emergent electronic states surviving to room temperature, for instance in indirect-gap heterostructures such as moiré superlattices between two individual TMD bilayers, where the valence band maximum is at Γ instead of K.

At Γ, the relevant low-energy orbitals arise purely from the WS$_2$ and WSe$_2$ layers, making any hybridization effects with the graphene layer negligible. However, near E$_F$, the effects of any interactions between the graphene bands and the underlying moiré superpotential should become evident. As shown in Fig. 3d, interactions between Dirac electrons in the graphene layer at K$_G$, and the moiré superlattice potential, give rise to multiple replica Dirac points which are separated by the WS$_2$/WSe$_2$ moiré wavevector (G$_{moiré}$ = 0.07 Å$^{-1}$), where even third-order replicas are evident. Due to the large number of replicas, a hexagonal arrangement can be clearly resolved. While an idealized triangular moiré superlattice should result in a perfect hexagon, we observed an elongation of this pattern along one axis by 21 ± 2 %, as shown in Fig. 4a. This dramatic distortion can, in fact, be generated by an uniaxial strain in only one of the layers relative to the other by a modest amount (0.7%), as illustrated in Fig. 4b (see supplementary information), consistent with reports from scanning tunneling microscopy (STM)[25]. Furthermore, the observation of a finite moiré wavevector rules out commensuration between WS$_2$/WSe$_2$ as the origin for the large splitting of the moiré minibands at Γ discussed in Fig. 3b-c.

There has been considerable debate as to whether observed replica Dirac cones in other systems such as graphene/SiC[26–28] or graphene/Ir (111)[29,30] might instead arise from final state diffraction of the photoelectrons. We believe that the replicas observed here are more likely from intrinsic interactions between the electrons in the graphene layer with the moiré superlattice potential. First, we do not observe replica Dirac cones in the g/WSe$_2$ heterostructure (Fig. 4c), as would also be expected if final-state photoelectron diffraction played a significant role. Second, the presence of energy gaps between the graphene and WSe$_2$ and WS$_2$ bands (as indicated by the white arrow in Fig. 3b) indicate strong interactions between the different layers. Although we cannot directly observe clear energy gaps between the different replica Dirac cones themselves (Fig. 4d), we suspect this is most likely due to the thermal broadening (T = 300 K) coupled with the experimental energy resolution (125 meV) being significantly larger than the strength of the moiré superpotential.

These clearly resolved moiré replicas suggest a new and unexplored pathway towards imprinting moiré superlattice potentials in layered materials. In conventional bilayer moiré superlattices, the size and orientation of the moiré wavevector is completely fixed by the twist angle and lattice mismatch between the two layers. On the other hand, for moiré superlattices imprinted onto a third layer, the moiré wavevector becomes an independently adjustable parameter set by the composition of the moiré bilayer, which does not depend on the third layer, and thus should allow for a much broader range of moiré superlattices and designer electronic materials which can be realized, such as long-wavelength superlattices in systems which do not have a close lattice match. In conventional moiré superlattices, the spatial variations in hybridization are known to be caused by changes in the local stacking between atoms[31]. On the other hand, the mechanism by which a moiré superlattice is imprinted onto a third material remains an important open question, and is unlikely to be due to spatial variations in local stacking, since in "imprinted" structures, different sites separated by a moiré wavelength may have very different local atomic structures and stacking. This suggests that other effects, such as spatially modulated charge transfer between $WS_2/WSe_2$ and graphene and topography, may be more relevant for determining the strength of the superlattice potential in imprinted moiré superlattices. Our observations have revealed unusually strong moiré superlattice effects that are inaccessible to the prevailing experimental tools and richer than previously thought, calling for a deeper theoretical understanding of these systems. In addition, this work suggests a new avenue of constructing superlattices with strong superpotentials and arbitrary moiré wavelengths.

**Methods:**

**Sample fabrication**

Atomically thin flakes were exfoliated from bulk crystals onto silicon substrates with 285 nm oxides. Monolayer $WSe_2$ and $WS_2$ were identified by their optical contrast. The crystal orientations of $WSe_2$ and $WS_2$ were determined by angle-resolved optical second-harmonic generation (SHG). A pulsed laser with duration of 100 fs at 800 nm was used as the excitation source and the cross-polarized SHG signal was collected at 400 nm. A six-fold symmetry of the SHG signal was resolved by rotating the excitation polarization with respect to the sample orientation. The heterostructure was assembled using a layer-by-layer dry transfer technique described in Ref. 32. The thin flakes were picked up, one by one, with a stamp consisting of a polycarbonate thin film attached to a polydimethylsiloxane. Monolayer graphene and few layer hBN were used as the top and bottom encapsulating layers, respectively. The zigzag direction of the $WSe_2$ and $WS_2$ monolayers were aligned by SHG. The completed heterostructure was released onto the $TiO_2$ conducting substrate by heating the stamp to 180 ºC. The heterostructure was then rinsed in chloroform and isopropyl alcohol to remove the melted polycarbonate film and cleaned using an atomic force microscope tip in the contact mode.

**ARPES measurements**

After assembly and cleaning, the samples were sealed in a vacuum chamber inside an inert gas glovebox, and then pumped to $10^{-5}$ torr before transportation and loading into the Microscopic and Electronic Structure Observatory (MAESTRO) UHV facility at beamline 7.0.2 of the Advanced Light Source in Berkeley. The samples were annealed at 300º C for 10 hours before ARPES measurements. ARPES data were collected using a hemispherical Scienta R4000 electron analyzer with the energy and momentum resolutions set to 125 meV and 0.015 A$^{-1}$, respectively. The photon energy was set to 70 eV with the incident beam focused using an x-ray capillary to a nominal spot size of ~1 μm. Samples were maintained at room temperature throughout measurements.

**DFT**

Density functional theory (DFT) calculations of freestanding graphene, WSe$_2$, and WS$_2$ were performed using the Perdew, Burke and Ernzerhof generalized gradient functional[33], as implemented in the Quantum ESPRESSO suite[34,35]. Full and scalar relativistic projector augmented wave[36] pseudopotentials from the PSlibrary[37] were used for WSe$_2$/WS$_2$ and graphene, respectively. For WSe$_2$ and WS$_2$ an energy cutoff of 120Ry (500Ry) was used for the wavefunction (charge density) and a Γ centered 16x16x3 Monkhorst–Pack (MP) mesh[38] was used for k-space discretization. For graphene an energy cutoff of 150Ry (800Ry) was used for the wavefunction (charge density) along with a gamma centered 21x21x5 MP mesh. The DFT calculations for WSe$_2$/WS$_2$ (graphene) were then downfolded onto a manifold of 22 (18) maximally localized Wannier functions[39,40] using the Wannier90 package[41] for interpolation onto the same k-point mesh as the ARPES data and for construction of the constant energy contours.

**References for Methods:**

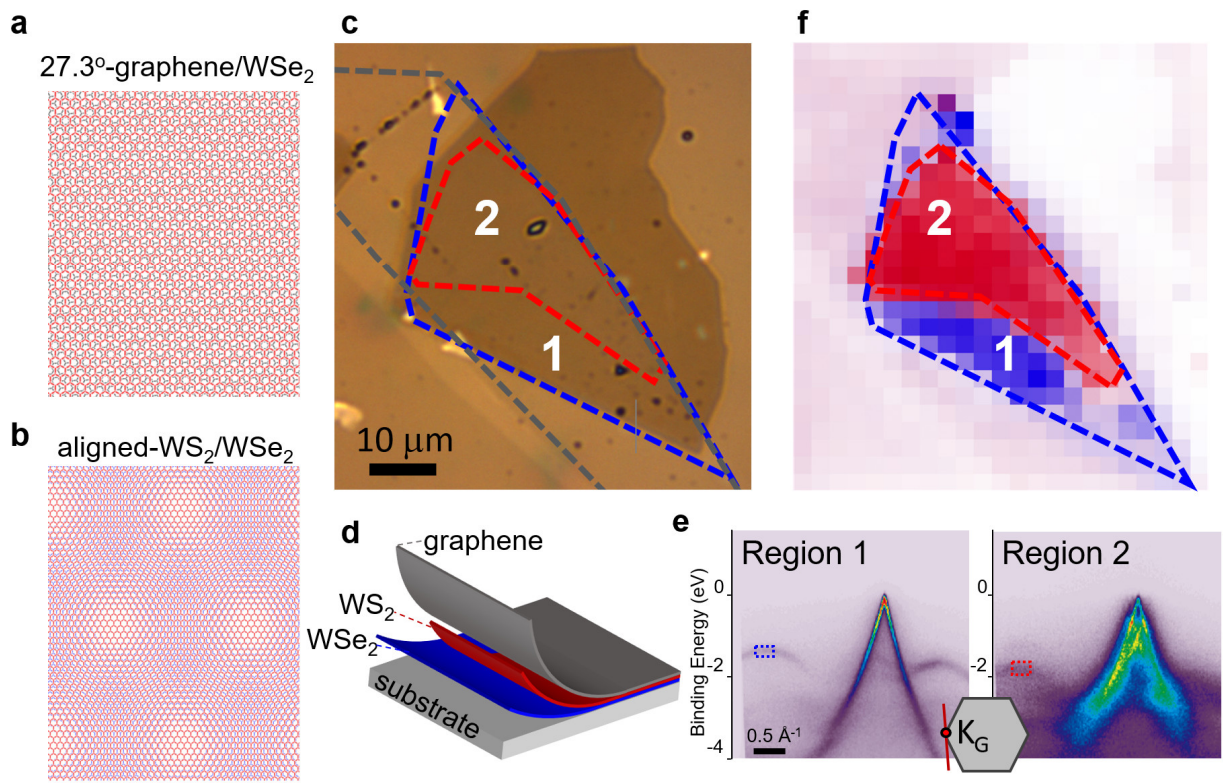

**Figure 1 | Moiré superlattices with wide-ranging wavelengths. a, b,** Schematic atomic arrangement of a 27.3°-graphene/WSe$_2$ (a) and aligned WS$_2$/WSe$_2$ (b) moiré superlattice. **c, d,** Optical micrograph (c) and schematics (d) of a moiré superlattice sample containing both g/WSe$_2$ (Region 1) and g/WS$_2$/WSe$_2$ (Region 2). **e,** ARPES spectra taken across the graphene Dirac point, for Region 1 and 2, respectively. **f,** A false-color spatial map of ARPES intensity by integrating the intensity inside the blue and red boxes shown in (e).

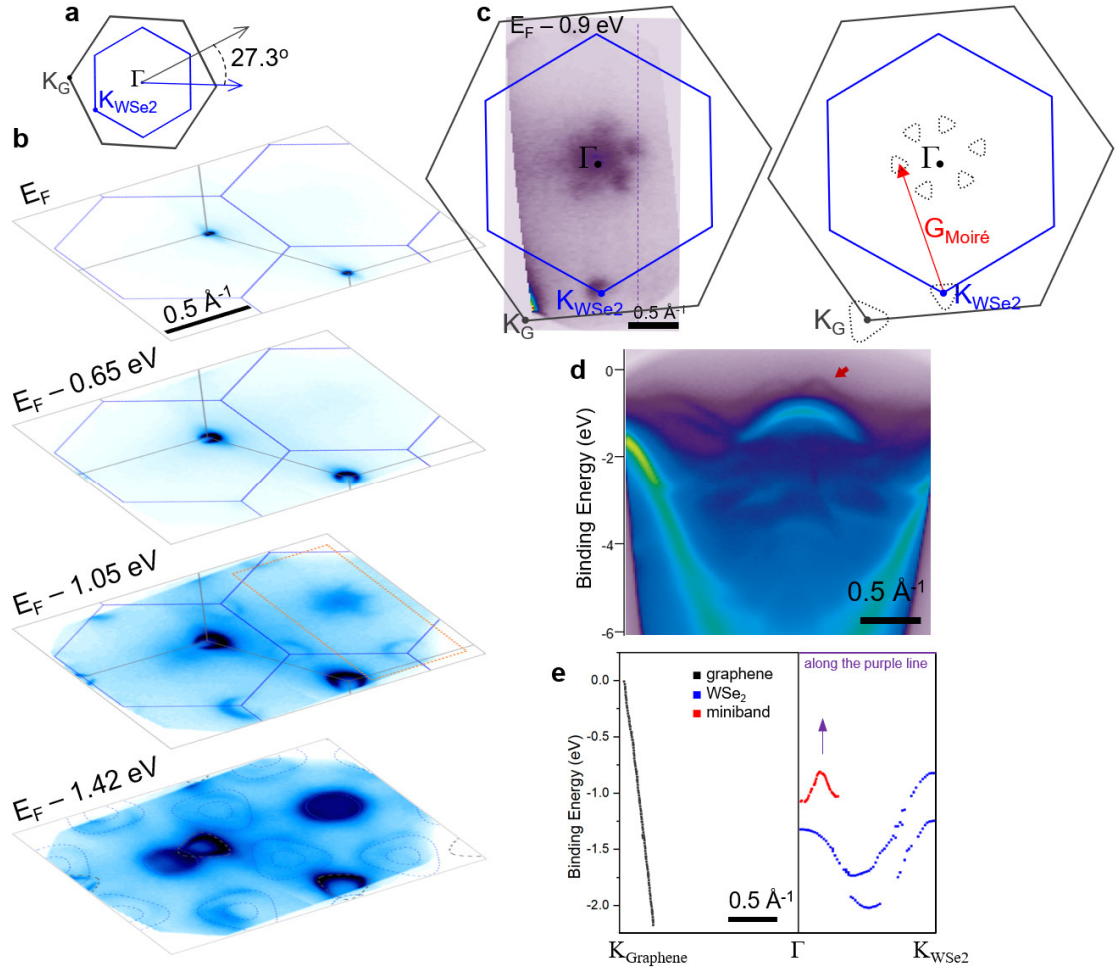

**Figure 2 | Direct observation of minibands in a 27.3°-graphene/WSe$_2$ moiré superlattice.**
**a,** Schematic of Brillouin zones of graphene (gray) and WSe$_2$ (blue). **b,** Constant energy contours at various binding energies. The gray and blue lines in the top three panels indicate the Brillouin zones of graphene and WSe$_2$, respectively. In the bottom panel, dashed lines show the bands produced by DFT calculations: gray for graphene and blue for WSe$_2$. **c,** Constant energy contour at $E_F - 0.9$ eV and the corresponding schematics showing the sextet of minibands at $K_{WSe2} + G_{moiré}$. **d,** ARPES cut along the purple dashed line in (c), where the miniband is indicated by a red arrow. **e,** Extracted dispersions of the miniband (red), WSe$_2$ valence bands along $\Gamma$-$K_{WSe2}$ (blue), and graphene Dirac cone (gray), showing the distinct dispersion of the minibands.

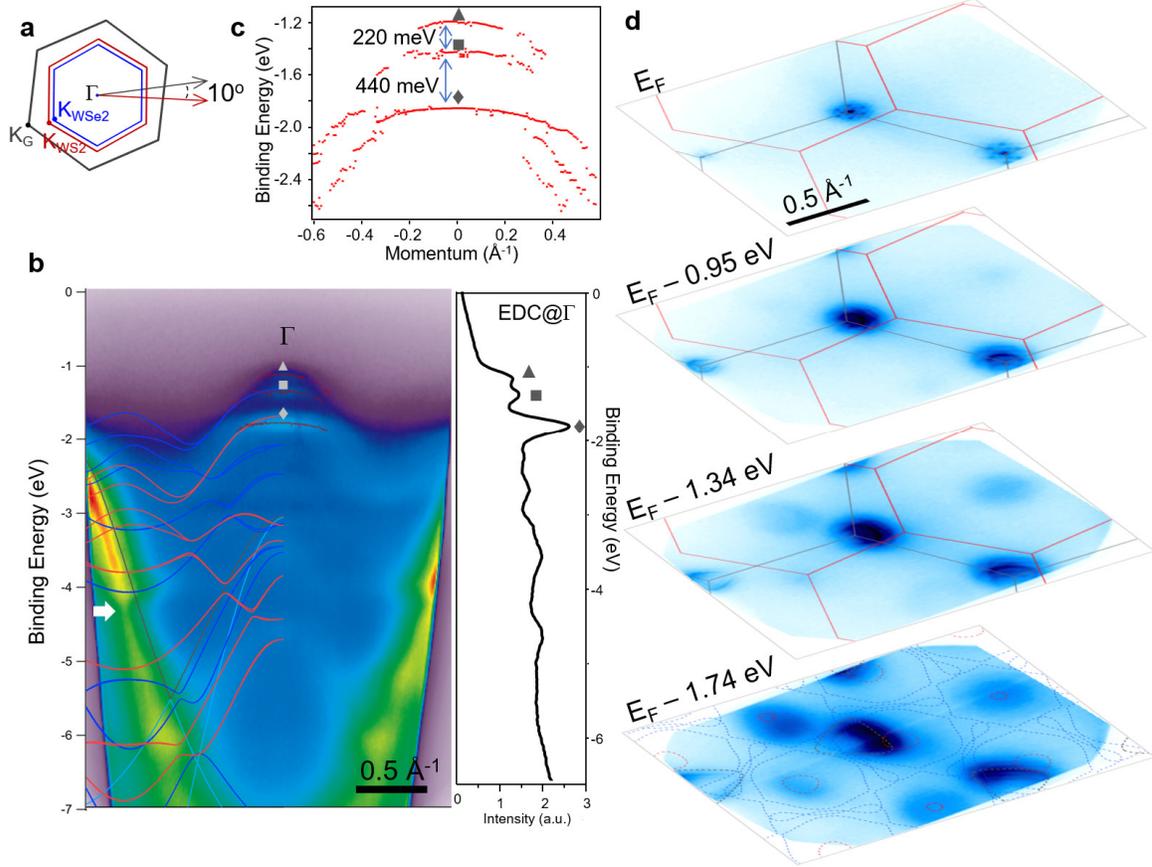

**Figure 3 │ Direct observation of minibands in a graphene/WS$_2$/WSe$_2$ moiré superlattice.**
**a,** Schematic of Brillouin zones of graphene (gray), WS$_2$ (red), and WSe$_2$ (blue). **b,** An ARPES cut across Γ point, where the left half is overlaid with the bands calculated by DFT (gray: graphene; red: WS$_2$; blue: WSe$_2$). Right panel: energy distribution curve (EDC) taken at Γ, and the moiré bands at different increasing binding energies are labeled with a triangle, a square, and a diamond, respectively. The fitted peak positions are further indicated with red dots in the ARPES cut. **c,** Extracted EDC peak positions showing the large energy splittings between the three bands observed at Γ. **d,** Constant energy contours at various binding energies. The gray and red lines in the top three panels indicate the Brillouin zones of graphene and WSe$_2$, respectively. In the bottom panel, dashed lines show the bands produced by DFT calculations: gray for graphene, red for WS$_2$, and blue for WSe$_2$.

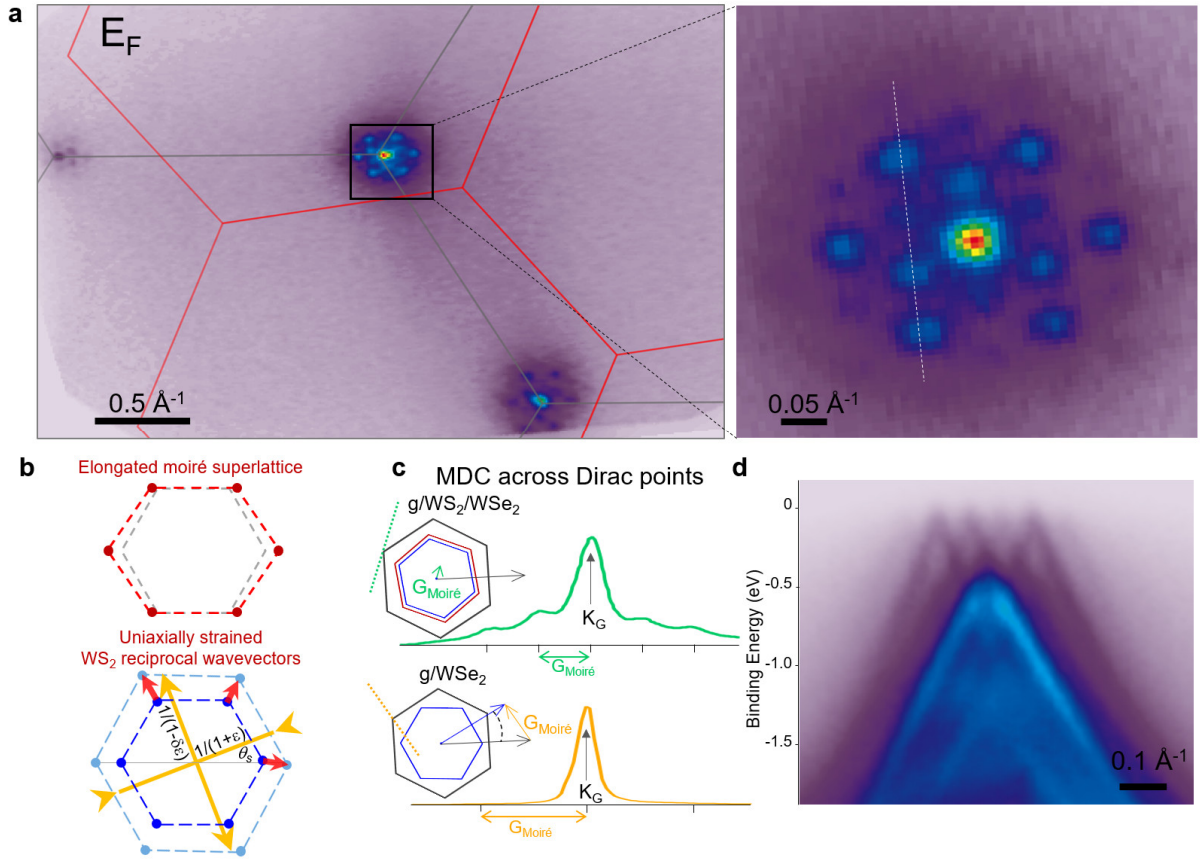

**Figure 4 | Imprinting effect of moiré superlattice in a graphene/WS$_2$/WSe$_2$ heterostructure. a,** Constant energy contours and an enlarged area near K$_G$ at E$_F$, showing multiple replica Dirac points separated by the WS$_2$/WSe$_2$ moiré wavevector. **b,** The elongated hexagon pattern of the observed replica Dirac points caused by a 0.7% uniaxial strain in the WS$_2$ layer (shown by its distorted reciprocal wavevectors of WS$_2$ in light blue, and the resulting moiré wavevectors in red). **c,** Momentum distribution curves (MDCs) taken across the Dirac point along the respective G$_{moiré}$ direction in g/WS$_2$/WSe$_2$ and g/WSe$_2$, as schematically shown by the dashed lines in the insets, and replica is only observed in the former heterostructure. **d,** An ARPES cut taken across multiple replica Dirac cones, along the white dashed line indicated in (a).

18